\documentclass[journal]{IEEEtran}
\usepackage{cite}
\usepackage[dvips]{graphicx}
\DeclareGraphicsExtensions{.eps}
\usepackage{amsmath}
\usepackage{mathptmx}
\usepackage{mathpazo} 
\usepackage{algorithmicx,algorithm}
\usepackage{algpseudocode}
\usepackage{array}
\usepackage[caption=false,font=footnotesize]{subfig}
\usepackage{stfloats}
\usepackage{diagbox}
\usepackage{multirow}
\usepackage{authblk}

\begin{document}
\title{ROI-Wise Material Decomposition in Spectral Photon-Counting CT}

\author{Bingqing~Xie,\thanks{B.Q. Xie, P. Niu and Y.M. Zhu  are with Univ Lyon, INSA Lyon, CNRS, Inserm, CREATIS UMR 5220, U1206, F-69621, Lyon, France.}
	~Pei~Niu,
	~Ting~Su,\thanks{T. Su, V. Kaftandjian and P. Duvauchelle  are with Univ Lyon, INSA Lyon, Laboratoire Vibrations Acoustique, F-69621 Villeurbanne, France.}
	~Val\'{e}rie~Kaftandjian,
	~Loic~Boussel,\thanks{L. Boussel and P. Douek are with Univ Lyon,
		CREATIS ; CNRS; Inserm; INSA-Lyon; Radiology Department, Hospices Civils de Lyon, Lyon, France.}
	~Philippe~Douek,
	~Feng~Yang,\thanks{F. Yang is with Beijing Jiaotong Univ, School of Computer and Information Technology, Beijing, China}
	~Philippe~Duvauchelle
	 and Yuemin~Zhu}

\maketitle

\begin{abstract}
Spectral photon-counting X-ray CT (sCT) opens up new possibilities for the quantitative measurement of materials in an object, compared to conventional energy-integrating CT or dual energy CT.
However, achieving reliable and accurate material decomposition in sCT is extremely challenging, due to  similarity between different basis materials, strong quantum noise and photon-counting detector limitations.
We propose a novel material decomposition method that works in a region-wise manner.
The method consists in optimizing basis materials based on spatio-energy segmentation of regions-of-interests (ROIs) in sCT images and performing a fine material decomposition involving optimized decomposition matrix and sparsity regularization.
The effectiveness of the proposed method was validated on both digital and physical data.
The results showed that the proposed ROI-wise material decomposition method presents clearly higher reliability and accuracy compared to common decomposition methods based on total variation (TV) or L1-norm (lasso) regularization.

\end{abstract}

\begin{IEEEkeywords}
X-ray CT, Material decomposition, Photon-counting detector.
\end{IEEEkeywords}

%

\section{Introduction}

\IEEEPARstart{S}{pectral} photon-counting X-ray computed tomography (sCT) is a new kind of multi-energy X-ray CT that offers new possibilities for getting insights into material components in an object, thanks to the advances in photon-counting detector (PCD) \cite{6412760,5740625,6748085}. Compared to conventional energy-integrating CT or dual energy CT, sCT can count the number of photons in multiple energy bins with one single exposure, $i.e.$ utilizing spectral information. This advantage enables efficient material decomposition that aims to quantitatively separate different materials present in a pixel.

Different approaches were developed to realize material decomposition: decompose projection data acquired at different energy bins into different material sinograms, each of which corresponds to a material ($i.e.$ the so-called basis material), based on their energy-dependent characteristics ($i.e.$ mass attenuation coefficients), and then reconstruct individually each spatial material image containing one single material (projection-domain approach) \cite{Schlomka2008,Zimmerman2015,Ducros2017}, or firstly reconstruct the spatial image from each energy bin sinogram and then decompose the reconstructed spatial images corresponding to different energy bins into spatial material images (image-domain approach) \cite{LeHuy2011,Taguchi2007,Li2015,Li2017}, or directly reconstruct spatial material images from projection data (one-step approach) \cite{FoygelBarber2016a,Mechlem2017,Hou2018}. 
The advantage of both projection-domain and one-step approaches is the more precise model that they directly decompose projection data (raw data) rather than after the procedure of reconstruction. The image-domain material decomposition has its strong point that it allows us to work directly on abundant morphological features. 

However, the reliability and accuracy of the above approaches are always impacted by unavoidable similarity between basis materials due to the similar dependence of mass attenuation coefficient curves on energy bins of different materials. Such similarity makes it difficult to separate the basis materials. The difficult basis material separation is further worsened by strong quantum noise. The latter also limits the ability of PCD to recognize photons between adjacent energy bins \cite{Schlomka2008}, which renders the basis material separation still more difficult.
Meanwhile, the performance of image-domain material decomposition is also susceptible to the quality of reconstructed sCT images including in particular the impact of beam hardening.

To improve the reliability and accuracy of material decomposition, 
it is important to make full use of more information beside the aforementioned spectral information \cite{Li2017,Zhang2017,Fadili2010}.
A straightforward way of realizing this is to exploit morphological information embedded in the reconstructed sCT images. 
Image-domain material decomposition is then an approach of choice.
In this paper, we investigate a novel image-domain material decomposition method 
by directly decreasing the impacts of  similarity between basis materials
 with the help of multiple features extracted from the reconstructed multi-energy spatial sCT images. 
The idea is to exploit the abundant information and high correlations in sCT images suffering from serious reconstruction errors and artifacts.
To do that, we perform basis material optimization by selecting basis materials according to their spatio-energy similarity in segmented region of interests (ROIs) of multi-energy sCT images, thus leading to so-called ROI-wise material decomposition.
To our knowledge, this is the first work to improve the mathematical condition of material decomposition through optimizing basis materials by means of spatio-energy segmentation.  

The rest of this paper is organized as follows. In Section \ref{sec.Method}, we describe the proposed method of ROI-wise material decomposition. Section \ref{sec.Experiments} presents experiments and results on both simulations and real data. 
Finally, Sections \ref{sec.Discussion} and \ref{sec.Conclusion} are respectively given discussion and conclusion.

\section{Model and method}\label{sec.Method}
This section firstly presents typical models of image-domain material decomposition in sCT. Then, the proposed ROI-wise material decomposition method is described in detail.

\subsection{Model of image-domain material decomposition}
In the model of image-domain material decomposition, spatial images should first be reconstructed.

Spatial images are reconstructed at each separated polychromatic energy bin in sCT. The mean measured signal (number of photons penetrating materials) recorded by a PCD for the $u$-th projection within the $i$-th energy bin can be modeled by:
\begin{equation}
\label{eq:2.1.1}
{{\bar s}_i}(u) = \int_ \mathbb{R}{{n_0}(E){d_i}(E) e^{-\int_{L(u)} {\mu} (\vec x,E){\rm d}{l}}{\rm d}E},
\end{equation}
where $n_0(E)$ is the spectral X-ray photon fluence, ${\mu} (\vec x,E)$ the linear attenuation coefficient at position or pixel $\vec x$ for energy $E$, $L(u)$ the $u$-th projection, and $d_i(E)$ the detector response function or bin sensitivity function describing the sensitivity of separating photons belonging to two adjacent energy bins. 
The measured signal is assumed to be corrupted by independent Poisson noise:
\begin{equation}
{{s}_i}(u) = \mathcal{P}(\lambda={{\bar s}_i}(u)),
\end{equation}
where $\mathcal{P}$($\lambda$) denotes the Poisson distribution of mean $\lambda$, and $s_i(u)$ the measured number of photons for the $u$-th projection in the $i$-th energy bin.
The aim of reconstruction is to obtain the linear attenuation coefficients at each energy bin:
\begin{align}
\label{eq:2.1.2}
&\mu(\vec x,i) = \mathop{\rm argmin}\limits_{{\mu}}\mathcal{D}(ln(\frac{{s_i}(u)}{\int_ \mathbb{R}n_0(E){\rm d}E}), \mu(\vec x,i))+\mathcal{R},
\end{align}
where $\mu(\vec x,i)$ denotes the reconstructed linear attenuation coefficients at the $i$-th energy bin, which also represents the reconstructed spatial image at the $i$-th energy bin, $\mathcal{D}$ the discrepancy function, and $\mathcal{R}$ the regularization term. 
Reconstruction in sCT suffers from severe noise problem because of the limited photon flux, as an immediate consequence of preventing the pileup effect of PCD which describes the distortion of recorded energy spectrum by coincident pulses \cite{Taguchi2010}.
Therefore, we solve the above reconstruction model with simultaneous algebraic reconstruction technique and total variation (SART-TV) algorithm that is commonly used for low-dose CT image reconstruction \cite{Chen2013}.

Once reconstructed, the multi-energy images are decomposed into the linear combination of mass attenuation coefficients weighted by the corresponding mass density, described by:
\begin{equation}
\label{eq:2.1.3}
{\mu} (\vec x,i) = \sum\limits_{\alpha  = 1}^{M} {\dot{\mu} _{m\alpha }}(i){{\rho _\alpha }(\vec x)},\quad i = 1,...,B,
\end{equation}
where ${\dot{\mu} _{m\alpha }}(i)$ designates the calculated effective mass attenuation coefficient  of the $\alpha$-th basis material at the $i$-th energy bin, $M$ the total number of basis materials, $B$ the total number of energy bins, and ${\rho _\alpha }(\vec x)$ the mass density of the $\alpha$-th basis material at pixel $\vec x$.
For clarity, material decomposition can be formulated in matrix form as:
\begin{align}
\label{eq:2.1.4}
&\boldsymbol{Y}=\boldsymbol{MX} + \boldsymbol{N},
\end{align}
where $\boldsymbol{N}$ denotes the noise and $\boldsymbol{M}$ the decomposition matrix, each column of which represents the effective mass attenuation coefficients of one basis material for the $B$ energies:
\begin{align}
\label{eq:2.1.5}
\quad \boldsymbol{M}=\left[{\begin{array}{*{20}{c}}
	{{\dot{\mu} _{m1}}(1)}& \ldots &{{\dot{\mu} _{mM}}(1)}\\
	\vdots & \ddots & \vdots \\
	{{\dot{\mu} _{m1}}(B)}& \cdots &{{\dot{\mu} _{mM}}(B)}
	\end{array}}\right].
	\end{align}
$\boldsymbol{Y} \in \mathcal{R}^{B\times N_P}$ and $\boldsymbol{X} \in \mathcal{R}^{M\times N_P}$ represent respectively the reconstructed multi-energy spatial images containing linear attenuation coefficients $\boldsymbol{\mu}$ and the mass densities $\boldsymbol{\rho}$ of basis materials with $N_p$ indicating the total number of pixels or voxels.
Theoretically, the decomposition matrix can be initialized by the effective mass attenuation coefficients, calculated using \cite{LeHuy2011}:
\begin{equation}
\label{eq:2.1.6}
\dot{\mu}_{m\alpha}(i)= \frac{{\int_{E\in E_i} {n_0 (E){{\mu} _{m\alpha}}(E)} dE}}{{\int_{E\in E_i} {n_0 (E)}}dE},\quad i = 1,...,B,
\end{equation}
where ${{\mu} _{m\alpha}}(E)$ is the theoretical mass attenuation coefficient at energy $E$ retrieved from NIST \cite{hubbell2015tables},  and ${\int_{E\in E_i} {n_0 (E)}dE}$ the total number of incident photons belonging to the $i$-th energy bin of width $E_i$. In other words, $\dot{\mu}_{m\alpha}(i)$ represents the averaged value of all the theoretical mass attenuation coefficients inside each single energy bin, which is an estimate of the true mass attenuation coefficient corresponding to that energy bin.

\subsection{ROI-wise material decomposition}
\begin{figure}[t]
	\centering
	\includegraphics[width=0.4\textwidth]{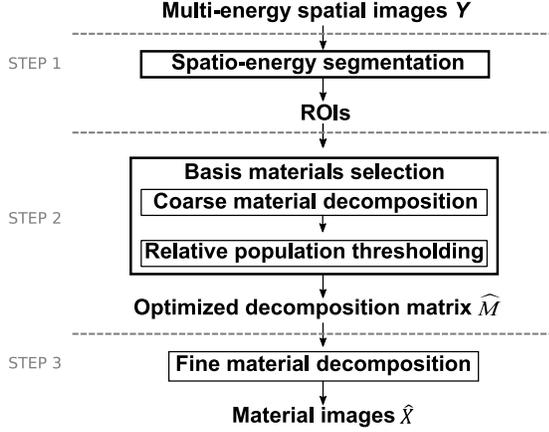}
	\caption{Diagram of the ROI-wise material decomposition algorithm. Tom to bottom: the spatio-energy segmentation based basis material optimization (blue) and noise-reduced composite image construction (green), and the fine material decomposition.}
	\label{figure.diagram}
\end{figure}

To make material decomposition more  reliable and accurate, we propose to fully exploit their spatio-energy similarity in
ROIs of multi-energy sCT images.
To do that, we perform a basis material optimization through reducing the impact of material similarity by means of spatio-energy segmentation.
The global diagram of the proposed ROI-wise material decomposition is illustrated in Fig. \ref{figure.diagram}.

Basis material optimization is first processed based on the spatio-energy segmentation that separates multi-energy spatial images into different ROIs and more details will be discussed in Section \ref{sec.D}. Next, the basis materials in $\boldsymbol{M}$ are selected by performing a coarse material decomposition followed by a relative population thresholding and details will be presented in Section \ref{sec.E}. In the third step, the basis material optimization result, namely the optimized decomposition matrix $\boldsymbol{\hat{M}}$, is finally utilized for the fine material decomposition.

%
%
%

\subsection{Spatio-energy segmentation}\label{sec.D}


We first obtain the ROIs of multi-energy images by spatio-energy segmentation, where each ROI represents a homogeneous area containing similar materials.
To achieve such spatio-energy segmentation,  spectral and morphological features of multi-energy images are used.

Spectral features are extracted by regrouping the sCT images  at all the energies as a single three-dimensional (3-D) image in which the energy is taken as a third dimension. 
Each pixel in the 3-D image has multi-energy values (energy-dependent $\mu$).
Pixels having different $\mu$ curves, $i.e.$ different spectral features, belong to different ROIs. 

Morphological features of sCT images are in fact energy-dependent because the characteristics of materials in sCT images are energy-dependent. As a result, structures imperceptible at certain energy bins may be distinguished more easily at another bin, depending on the physical characteristics of materials.
Therefore,  we take the morphological features from the energy bin having the most reliable segmentation by evaluating a pre-processing segmentation based on Gaussian mixture model (GMM) at each single energy bin.
Then, the extracted structures are treated as the common constraint for images at all energy bins.
The spectral and morphological features are then exploited jointly for segmentation based on clustering.


We use the kernel k-means method for the fusion of spectral and morphological features. 
The main advantage of kernel k-means is that it can make full use of kernel properties, which provides the ability to combine different features \cite{Dhillon2004,Fauvel2012}. 
More precisely, we utilize the kernel k-means method to automatically segment the pixels in the 3-D image:
\begin{align}
\begin{small}
\label{eq:2.3.1} \mathop{\rm argmin}\limits_{\boldsymbol{m}_k} \sum\limits_{k=1}^K \sum\limits_{\boldsymbol{y}\in{\pi}_k}\lVert \Phi(\boldsymbol{y})-\boldsymbol{m}_k\rVert^2_2,\quad s.t. \ \boldsymbol{m}_k = \frac{\Phi(\boldsymbol{y})}{\lVert \pi_k\rVert_1},
\end{small}
\end{align}
where $\Phi$ represents a non-linear transform function, $\pi_k$ a partitioning of multi-energy pixel values $\boldsymbol{y}\in \mathcal{R}^B$, $K$ the total number of clusters ($i.e.$ the total number of ROIs), $\boldsymbol{m}_k$ the ideal cluster values, and $\lVert\cdot\rVert_2$ and $\lVert\cdot\rVert_1$ denote L-2 and L-1 norm, respectively. As an enhanced algorithm of normal k-means, kernel k-means can separate vectors in a high-dimensional feature space based on the non-linear transform. 
The non-linear transform can be calculated with a convenient kernel, and new kernel can be constructed by linearly adding two basic kernels, denoted by $\boldsymbol{\mathcal{K}}_1$ and $\boldsymbol{\mathcal{K}}_2$ respectively, $i.e.$ $\theta_1\boldsymbol{\mathcal{K}}_1+\theta_2\boldsymbol{\mathcal{K}}_2$, which is still a kernel. We utilize this property to combine spectral and spatial features which are separately selected by two kernels, named as $\boldsymbol{\mathcal{K}}^{spectrum}$, $\boldsymbol{\mathcal{K}}^{space}$. The new kernel is then obtained:
\begin{align}
\begin{small}
\label{eq:2.3.2}
\boldsymbol{\mathcal{K}}_{\sigma,\lambda} = (1-\theta)\boldsymbol{\mathcal{K}}^{spectrum}_{\sigma} + \theta\boldsymbol{\mathcal{K}}^{space}_{\sigma},
\end{small}
\end{align} 
where $0\le\theta\le1$ and $1-\theta$ designate the weights for spatial and spectral kernels, respectively. $\theta$ controls the relative impacts of spatial and spectral features on the final kernel.

More precisely, the kernel for spectral features is a Gaussian radial basis kernel:
\begin{align}
\begin{small}
\label{eq:2.3.3}
(\boldsymbol{\mathcal{K}}^{spectrum}_{\sigma})_{ij} = \kappa_\sigma(\boldsymbol{y}_i,\boldsymbol{y}_j)
= {\rm exp} (-\frac{\lVert \boldsymbol{y}_i - \boldsymbol{y}_j \rVert_2^2}{2 {\sigma}^2}),
\end{small}
\end{align} 
where $\kappa$ denotes the Gaussian radial basis kernel and $\sigma$ the variance of the corresponding Gaussian distribution.
The kernel for spatial features is also a Gaussian radial basis kernel:
\begin{align}
\begin{small}
\label{eq:2.3.4}
(\boldsymbol{\mathcal{K}}^{space}_{\sigma})_{ij} = \kappa_\sigma(\boldsymbol{y}_{si},\boldsymbol{y}_{sj}) = {\rm exp} (-\frac{\lVert \boldsymbol{y}_{si} - \boldsymbol{y}_{sj} \rVert_2^2}{2 {\sigma}^2}).
 \end{small}
\end{align}
However, the pixel values $\boldsymbol{y}_{si} \in \mathcal{R}^B$ calculated in $\boldsymbol{\mathcal{K}}^{space}_{\sigma}$ are those of the pre-processed 3-D image $\boldsymbol{Y}_s$ containing the 'labels' of morphological information. The pre-processing aims to detect spatial features from the image having the most reliable morphological information. In practice, we firstly divide pixels in each spatial image by a classical classification method involving GMM, which is a probabilistic model that assumes that all the data points are generated from a mixture of a limited number of Gaussian distributions:
\begin{align}
\label{eq:2.3.5}
&p_M({y}^b)=\sum\limits_{i=1}^{K}\alpha_ip(y^b|{m}_i^G,{\sigma}_{i}),
\end{align} 
where $p(y^b|{m}_k^G,{\sigma}_{i})$ represents the $i$-th Gaussian distribution with means ${m}_i^G$ and covariance ${\sigma_{i}}$, $\alpha_i$ the corresponding weights, $K$ the total number of components, and $y^b$ the $b$-th bin value of pixel $\boldsymbol{y}$ in the 3-D image. 
Then, the spatial image at each energy bin is segmented into $K$ areas by classification.
In our case the probability density function $p_M({y}^b)$ of GMM is used to estimate the reliability of  segmentation results. The reliability of segmentation is evaluated by the optimal value of the loglikelihood cost function:
\begin{align}
\begin{small}
\label{eq:2.3.6}
LL(\boldsymbol{Y}^b)={\rm ln}\left(\prod\limits_{i=1}^{N_P} p_M(y_{i}^b)\right),
\end{small}
\end{align} 
where $N_P$ denotes the total number of pixels in the two-dimensional (2-D) image at each bin.
The segmentation at certain energy bin with larger loglikelihood value is believed to be more reliable, and then the detected edges in the image at that energy bin are taken as the common edges for all the spatial images at different bins. Note that the images at different bins should be normalized before  comparing loglikelihood values. 
After that, the mean value of pixels inside each segmented area at each bin is calculated:
\begin{align}
\begin{small}
{\bar{y}^b_{\{\pi_k\}}}=\frac{1}{N_k}\sum_{i=1}^{N_k}y^b_i, \quad s.t. \quad \ y^b_i\in {\{\pi_k\}},
 \end{small}
\end{align}
where $y^b_i\in {\{\pi_k\}}$ represents the $i$-th pixel value in the $k$-th segmented area at the $b$-th bin, and $N_k$ the total number of pixels insider the $k$-th area. 
Then, a new 3-D image $\boldsymbol{Y}_s$ is produced by assigning its pixel value the corresponding mean value ${\bar{y}^b_{\{\pi_k\}}}$:
\begin{align}
&y^b_{si} = {\bar{y}^b_{\{\pi_k\}}}, \quad s.t. \quad \ y^b_{si}\in {\{\pi_k\}},
\end{align}
where $y^b_{si}\in {\{\pi_k\}}$ represents the $i$-th pixel value of $\boldsymbol{Y}_s$ in the $k$-th segmented area at the $b$-th bin.
In other words, a 'label' is attributed to each pixel associated with its spatial features. Thus, each multi-dimensional pixel is now assigned to two different values: the original value ($\boldsymbol{y}$) containing spectral feature and the mean value ($\boldsymbol{y}_s$) containing morphological feature, of which the features can be extracted by different kernels and fused together by our final kernel given by (\ref{eq:2.3.2}).


\subsection{Basis materials Selection}\label{sec.E}
The second step of basis material optimization is to update the initialized decomposition matrix under the principle of keeping the minimum needed number of basis materials (with respect to ground-truth).
To do this, we introduce a coarse material decomposition in each ROI, which exploits the sparse nature by L-1 norm ($lasso$ \cite{Tibshirani1996}):
\begin{align}
\begin{small}
\label{eq:2.3.7}\mathop{\rm argmin}\limits_{\boldsymbol{X}}\frac{1}{2}\lVert \boldsymbol{Y}-\boldsymbol{MX}\rVert_F^2+\lambda \lVert \boldsymbol{X}\rVert_1,
\end{small}
\end{align} 
where $\frac{1}{2}\lVert \boldsymbol{Y}-\boldsymbol{MX}\rVert_F^2$ is the data fidelity term calculated in terms of Frobenius Norm $\lVert\cdot\rVert_F$ and $\lambda$ the Lagrange multiplier.
The coarse material decomposition method is subject to obvious unreliability of detecting materials and poor accuracy \cite{Xie2019}. It can nevertheless help indicate us which basis materials should be considered nonexistent, while the others are more plausible under certain selection criterion. Therefore, we propose a relative population thresholding (RPT) method to determine which basis material deserves to be selected. 
The RPT method is based on the population percentage defined as the ratio of the number of pixels containing a decomposed material $N_{x}$ to the total number of pixels $N_{ROI}$ in each ROI. Only the materials with percentage above a threshold  will be preserved in each ROI. 
The physical meaning of RPT is partly involved with the aforementioned local property in both spectral and morphological domains. Local property limits the distribution of basis materials, which means that a corresponding minimum population percentage for all the ROIs should exist.
In other words, the threshold of population percentage is associated with the aggregation degree of materials.
After basis material optimization, we obtain a new $\boldsymbol{\hat{M}}$ from $\boldsymbol{M}$ in the $k$-th ROI, as:
\begin{align}  
\label{eq:2.3.8}\nonumber&
\begin{small}
\boldsymbol{\hat{M}}_k \leftarrow (\boldsymbol{M})_{\alpha},
\end{small} \\
&
\begin{small}
s.t. \ \frac{N_{{x}{\alpha}k}}{N_{ROIk}} \geq T,
\end{small} 
\end{align}
where $\boldsymbol{\hat{M}}_k\in \mathcal{R}^{B\times M'}$ denotes the optimized decomposition matrix containing $M'$ entries in the $k$-th ROI, $(\boldsymbol{M})_{\alpha}$ the $\alpha$-th basis material in decomposition matrix $\boldsymbol{M}$, $\leftarrow$ the operator of assigning a material from ${\boldsymbol{M}}$ to $\hat{\boldsymbol{M}}$, $N_{{x}{\alpha}k}$  the number of pixels containing the $\alpha$-th basis material in the $k$-th ROI, and ${T}$ the population threshold.

%
%
%
\subsection{Fine material decomposition}
Fine material decomposition is the last step in the proposed ROI-wise material decomposition method, which involves two terms: data fidelity term based on the optimized decomposition matrix $\hat{\boldsymbol{M}}$ and sparsity regularization term. 
%

Mathematically, we formulate the fine material decomposition as:
\begin{align}
\label{eq:2.4.1}
\nonumber&
\begin{small}
\mathop{\rm argmin}\limits_{\boldsymbol{X}}\sum\limits_{k} \frac{1}{2}\lVert \boldsymbol{Y}_{k}-\boldsymbol{\hat{M}}_{k}\boldsymbol{X}_k\rVert_F^2+\lambda\lVert\boldsymbol{X}_{k}\rVert_1
\end{small}\\
&
\begin{small}
= \mathop{\rm argmin}\limits_{\boldsymbol{X}} \frac{1}{2}\lVert \boldsymbol{Y}-\boldsymbol{\hat{M}X}\rVert_F^2+\lambda\lVert\boldsymbol{X}\rVert_1,
\end{small}
\end{align} 
where $\lVert \boldsymbol{Y}-\boldsymbol{\hat{M}X}\rVert_F^2$ denotes the data fidelity term.
The above fine material decomposition model is solved via the alternating direction method of multipliers (ADMM) iteration method \cite{Boyd2010, Jiao2016a}.

\section{Experiments and results}\label{sec.Experiments}
The performance of the proposed ROI-wise material decomposition method was evaluated on both digital and physical phantom data. 
\subsection{Digital phantom data}
\subsubsection{Digital phantom data generation}
\begin{figure}
	\centering
\subfloat[]{\label{fig.DigitalPhantom.a}
	\includegraphics[width=0.25\textwidth]{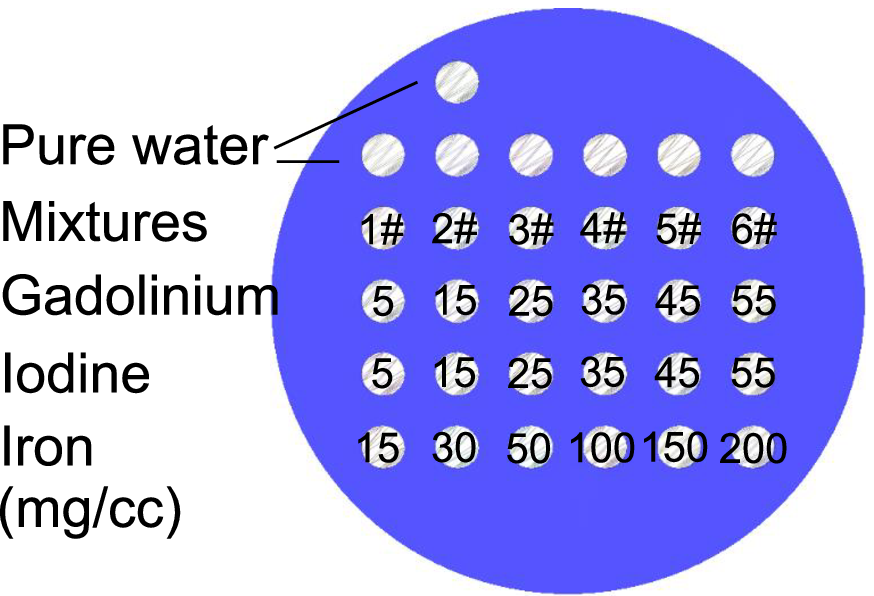}}	\
\subfloat[]{\label{fig.DigitalPhantom.b}
	\includegraphics[width=0.2\textwidth]{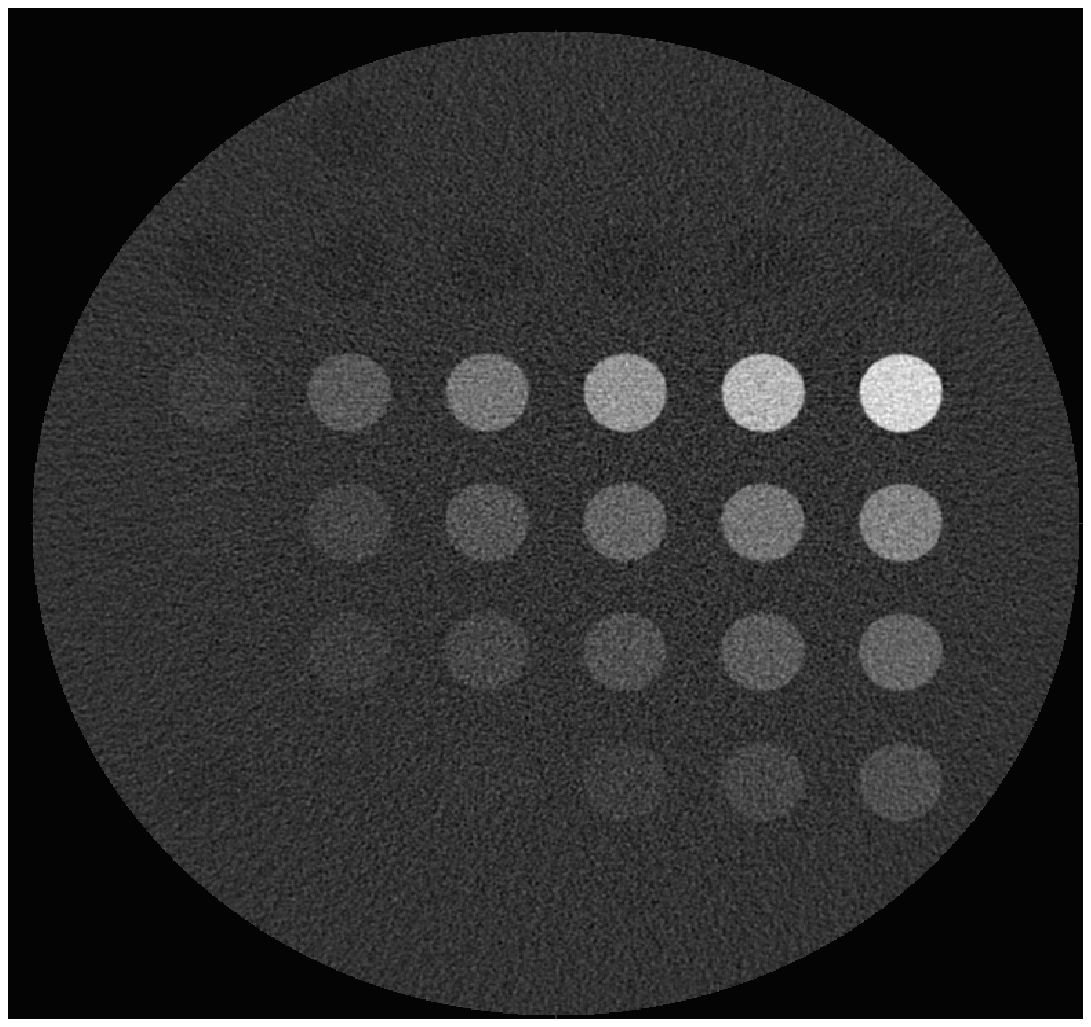}}	
\caption{(a) Digital phantom; (b) the reconstructed image of digital phantom at the first energy bin.}
\end{figure}

The data of sCT was simulated using the software Virtual X-ray Imaging (VXI) \cite{Duvauchelle} with a detector of 700 pixels by 1200 views over $360^{\circ}$. The X-ray energy bins were set as: 30-40 keV, 40-50 keV, 50-60 keV, 60-70 keV and 70-80 keV. The reconstructed phantom (one slice) has 780*780 pixels and contains five basis materials: water, PMMA, gadolinium (Gd), iodine (I) and iron (Fe), as shown in Fig. \ref{fig.DigitalPhantom.a}. The number on each disk indicates the concentration of materials (mg/cc). 
Note that the disk with '$\#$' stands for mixture inserts that contain three basis materials (gadolinium, iodine and iron) with the same concentration in each column. The reconstructed image of the first bin is shown in Fig. \ref{fig.DigitalPhantom.b}. The mass attenuation coefficients were retrieved from NIST \cite{hubbell2015tables}.

\subsubsection{Image quality metrics}
The performance of the proposed material decomposition was quantitatively evaluated using different metrics. The normalized Euclidean distance was utilized for the accuracy \cite{Ducros2017}, considering its better evaluation for various concentrations compared to the common metric mean squared error. Given both the $m$-th decomposed basis materials $\boldsymbol{x}_m$ and  ground-truth $\boldsymbol{x}_m^{gt}$, the normalized Euclidean distance is:
\begin{align}
\begin{small}
error_{m}=\frac{\lVert\boldsymbol{x}_m - \boldsymbol{x}_m^{gt}\rVert_2}{\lVert \boldsymbol{x}_m^{gt}\rVert_2}.
\end{small}
\end{align} 
The smaller the normalized Euclidean distance, the more accurate the decomposition precision.
To evaluate the reliability of material decomposition, we introduced two criteria: false positive (FP) and false negative (FN).
The FP or FN rate is calculated as the ratio of the number of  wrongly recognized pixels $N_{FP}$ ($i.e.$ for materials inexistent in  ground-truth but in decomposition results) or unrecognized $N_{FN}$ ($i.e.$ for materials existent in ground-truth but not decomposed in results) to the total number of pixels in all the ROIs $N_{ROIs}$:
\begin{align}
\begin{small}
FP = \frac{N_{FP}}{N_{ROIs}};\quad
FN = \frac{N_{FN}}{N_{ROIs}}.
\end{small}
\end{align} 
For sCT material decomposition, a smaller FP rate means smaller errors of confusing different materials, while a larger FN occurs when existing materials cannot be recognized. In other words, the smaller the FP and FN, the more reliable the material decomposition. 

\subsubsection{Results}
Two other methods were compared to the proposed ROI-wise material decomposition method: (a) the common TV method; (b) the Coarse method in (\ref{eq:2.3.7}). Note that the 'Coarse' method is also an intermediate step of the ROI-wise method, which can be utilized to evaluate the impacts of the involved  basis material optimization and energy averaging-based denoising. For clarity, we denote the proposed ROI-wise material decomposition by 'ROI' method in the figures.

\begin{figure}
	\centering
	{\includegraphics[width=0.45\textwidth]{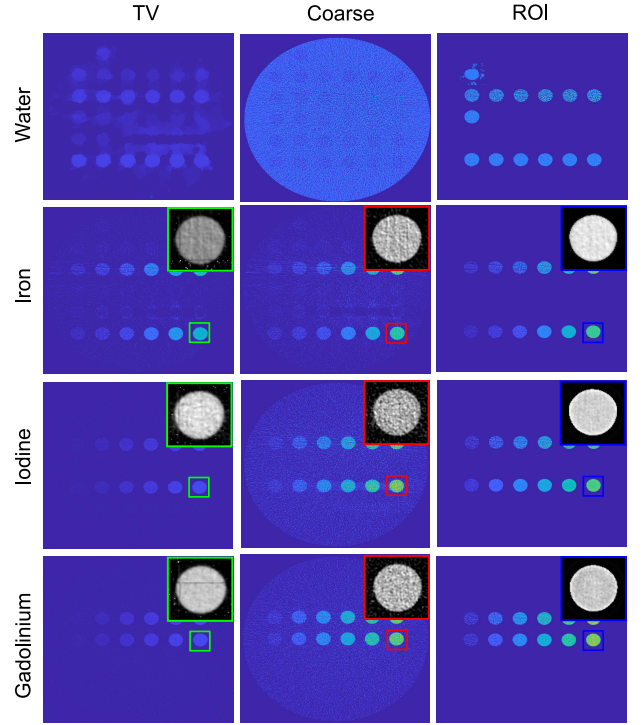}}
	\caption{The results of material decomposition using three methods on digital phantom. Left to right: TV, Coarse and ROI decompositions.}
	\label{fig.result.DP1}
\end{figure}

The decomposition results of four basis materials (water, iron, iodine and gadolinium) are shown in Fig. \ref{fig.result.DP1}. In contrast to the other two methods, the proposed ROI-wise material decomposition method shows better detection ability in terms of  edge-preserving performance. 

Visually, there are obviously more errors for water-like materials than for iron, iodine and gadolinium. 
Fortunately in practical applications, the quantitative information of the last three materials are more useful, as a result of which we will pay more attention to the materials of interest: iron, iodine and gadolinium.
As an illustration, we list $error_{m}$ of iron, iodine and gadolinium for the three methods in Table \ref{Tab.error.tot.sim}. Clearly, ROI-wise method has the smallest $error_{m}$ for all the three materials  compared to TV or Coarse method.

\begin{table}

	\centering
	\caption{The normalized Euclidean distance $error_{m}$ of different decomposition methods on digital phantom.}
	\begin{tabular}{c||ccc}	
		\hline
		\diagbox{Materials}{Methods}&TV&Coarse&ROI\\
		\hline
		\hline
		Iron&		0.21&0.23&0.14  \\
		Iodine&		0.28&0.37&0.15	\\
		Gadolinium&	0.26&0.31&0.12	\\	
		\hline		
	\end{tabular}
	\label{Tab.error.tot.sim}

\end{table}

The results in terms of FP and FN are given in Table \ref{Tab.FNFP.simu}. The proposed ROI-wise material decomposition method has a much smaller FP compared to the TV and Coarse methods. In terms of FN, the three methods exhibit similar performance for iodine and gadolinium, but the ROI-wise method leads to larger errors for iron.

\begin{table}
	\centering
	\caption{The FP and FN rates of different decomposition methods on digital phantom.}
	\begin{tabular}{cc||ccc}
		\hline
		\multicolumn{2}{c||}{\diagbox{Materials}{Methods}}&TV&Coarse&ROI\\
		\hline
		\hline
		Iron      &\multirow{3}{*}{FP (\%)}   &13.6&18.4&0.02\\
		Iodine    &     				      &5.2 &9.6 &0.01\\
		Gadolinium&						      &4.9 &6.1 &0.01\\
		\hline
		Iron	  &\multirow{3}{*}{FN (\%)}   &9.1 &14.8&19.5\\
		Iodine    &    						  &29.3&28.5&29.9\\
		Gadolinium& 						  &28.0&26.3&30.3\\		
		\hline
	\end{tabular}
	\label{Tab.FNFP.simu}
\end{table}

\subsection{Physical phantom data}
\subsubsection{Physical phantom data acquisitions}
\begin{figure}
	\centering
	\subfloat[]{\label{fig.PhysicalPhantom.a}
		\includegraphics[width=0.25\textwidth]{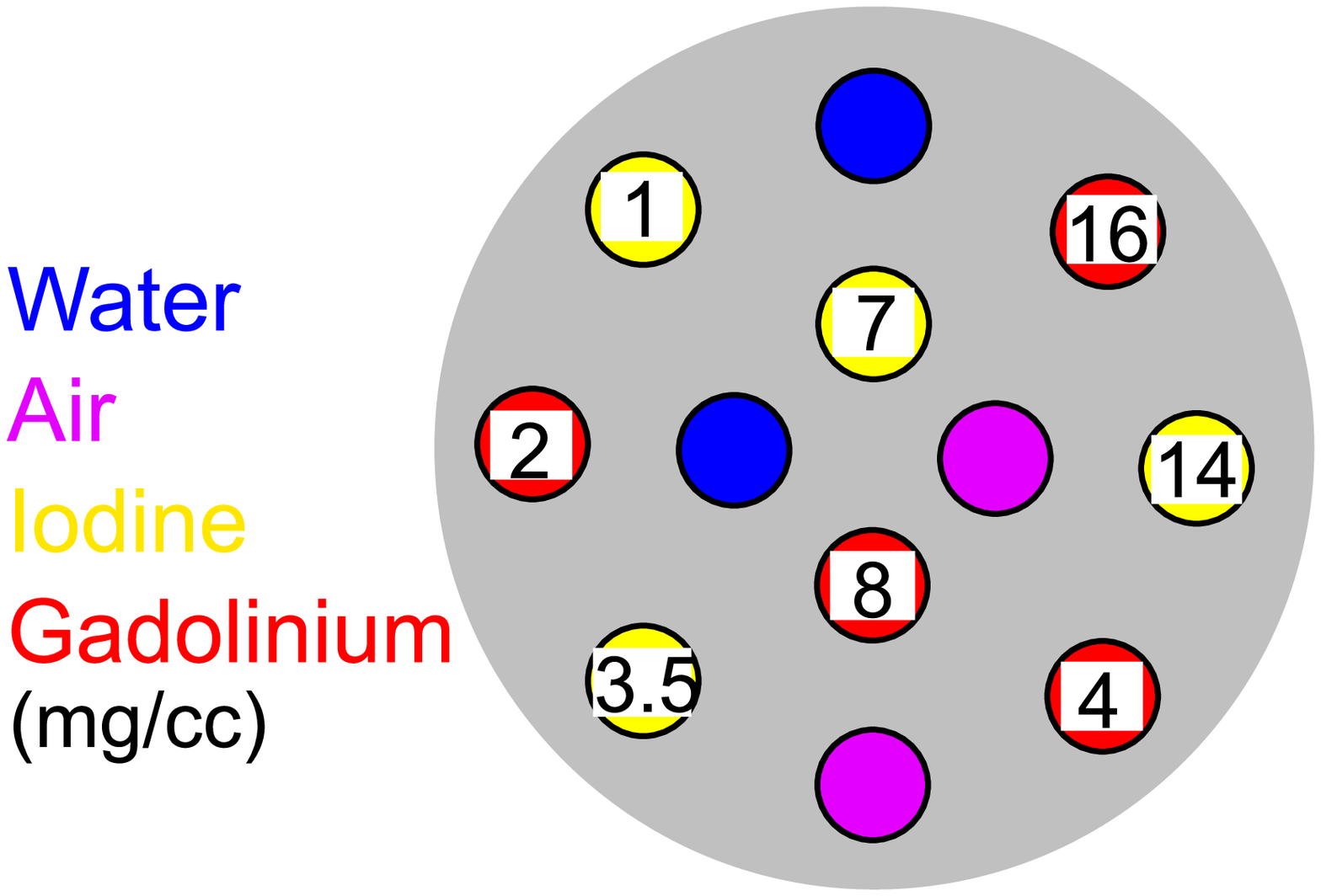}}	\
	\subfloat[]{\label{fig.PhysicalPhantom.b}
		\includegraphics[width=0.2\textwidth]{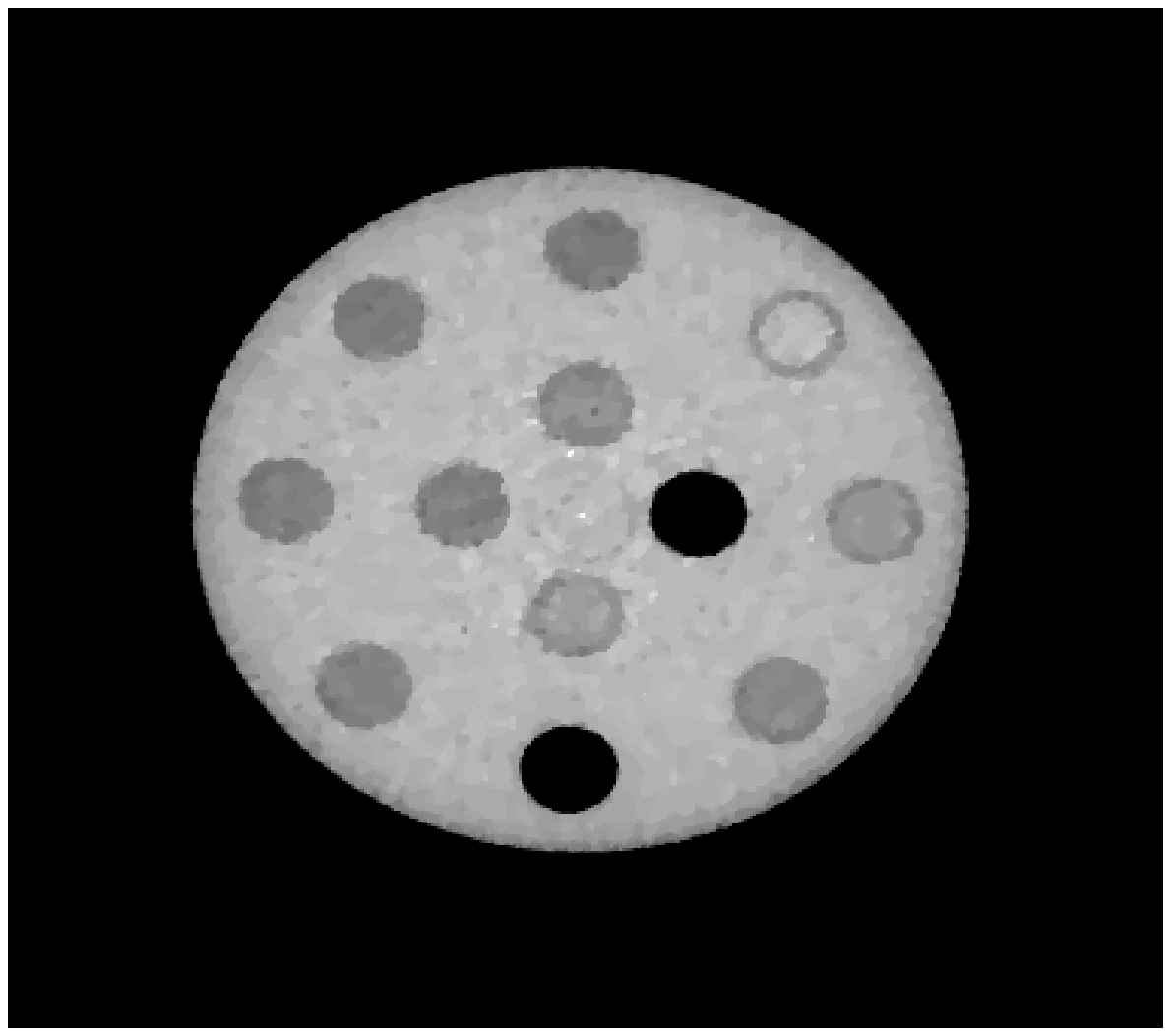}}	
	\caption{(a) Physical phantom; (b) the reconstructed image of physical phantom at the first energy bin.}
\end{figure}

The physical phantom was acquired on a Philips sCT prototype \cite{Si-Mohamed,Cormode}. The scan consists of 2400 projections; each projection has 643 parallel rays; each ray contains 5 energy bins: 30-50 keV, 51-61 keV, 62-71 keV, 72-82 keV and 83-130 keV. The incident photons $n_0(E)$ and detector bin response function $d_i(E)$  were provided by the manufacturer of the sCT prototype. The physical phantom is shown in Fig. \ref{fig.PhysicalPhantom.a}. The annotation is the same as for digital phantom. We reconstructed the spatial image at each energy bin using SART-TV, and the reconstructed image of the first bin is shown in Fig. \ref{fig.PhysicalPhantom.b}, where ring artifacts are obvious. The image reconstructed at each bin has a size of 380*380.

\subsubsection{Results}
\begin{figure}
	\centering
	\includegraphics[width=0.45\textwidth]{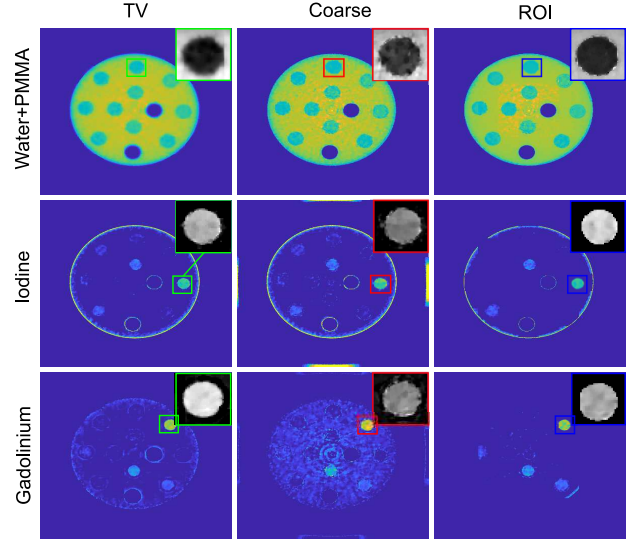}
	\caption{The results of material decomposition on  physical phantom using three methods. Left to right: TV, Coarse and ROI decompositions.}
	\label{figure.decomposition.gray}
\end{figure}
\begin{figure}
	\centering
	{\includegraphics[width=0.4 \textwidth]{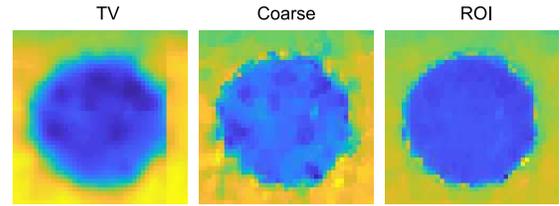}}
	\caption{The results of decomposed water-like materials in selected areas in Fig. \ref{figure.decomposition.gray}.}
	\label{figure.PP1}
\end{figure}

The results of material decomposition on the physical phantom are shown in Fig. \ref{figure.decomposition.gray}.
The proposed ROI-wise material decomposition method has clearly better performance in detection and quantification compared to TV and Coarse methods. 
Firstly, our method gives much better morphological accuracy, even for water-like materials. The selected areas for water-like materials using both TV and ROI-wise methods, shown in Fig. \ref{figure.PP1}, illustrate that the edges of the selected areas are severely blurred by TV method and mosaicked by Coarse method.
In contrast, the edges are substantially better preserved by ROI-wise method. 
As observed, none of the three methods has accurately decomposed iodine of concentration 1 $mg/cc$. For example, ROI-wise method was not able to recognize iodine inside the disk, while the other two methods were not able to separate the iodine from water or gadolinium of 2 $mg/cc$.

Secondly, more quantitatively as shown in Table \ref{Tab.error.relative}, ROI-wise method has clearly smaller normalized Euclidean distance. Specially, ROI-wise method has higher accuracy for gadolinium than for iodine, which is the same result as in the digital phantom case.

\begin{table}
	\centering
	\caption{The normalized Euclidean distances $error_{m}$ of different decomposition methods on physical phantom.}
	\begin{tabular}{c||ccc}
		\hline
		\diagbox{Materials}{Methods}&TV&Coarse&ROI\\
		\hline
		\hline
		Iodine&		0.41&0.46&0.38	\\
		Gadolinium&	0.26&0.98&0.23	\\	
		\hline
	\end{tabular}
	\label{Tab.error.relative}
\end{table}

Table \ref{Tab.FNFP} lists both FN and FP for different materials using the three methods among which the proposed ROI-wise method produces the smallest FP rate. The proposed ROI-wise method has 64\% FP rate improvement for iodine and 96\% for gadolinium compared to  Coarse method. In terms of FN, ROI-wise method leads to obvious errors especially for iodine.

\begin{table}
	\centering
	\caption{The FP and FN rates of different decomposition methods on physical phantom.}
	\begin{tabular}{cc||ccc}
		\hline
		\multicolumn{2}{c||}{\diagbox{Materials}{Methods}}&TV&Coarse&ROI\\
		\hline
		\hline
		Iodine    &\multirow{2}{*}{FP (\%)}   &16.7&21.2&7.7 \\
		Gadolinium&						      &31.3&78.5&3.5 \\
		\hline
		Iodine    &\multirow{2}{*}{FN (\%)}   &3.9 &2.8 &25.7 \\
		Gadolinium& 						  &3.9 &1.4 &8.0 \\		
		\hline
	\end{tabular}
	\label{Tab.FNFP}
\end{table}

Concerning the influence of important kernel parameters including the weight of spatial kernel $\theta$ and the variance of the Gaussian distribution $\sigma$, the decomposition performance of iodine with different [$\theta$ $\sigma^2$] is listed in Table \ref{Tab.error.parameters} (noting that gadolinium has the same trend for [$\theta$ $\sigma^2$], no longer listed here).
A too small or big $\theta$ can lead to larger  normalized Euclidean distance $error_{m}$. An intermediate value of $\theta= 0.2$ was selected in the experiment. 
 $\sigma$ shows less impact on the decomposition, as a result of which we chose $\sigma^2=0.5$ according to its overall smaller $error_m$ for different $\theta$. 

\begin{figure}
	\centering
	\includegraphics[width=0.25\textwidth]{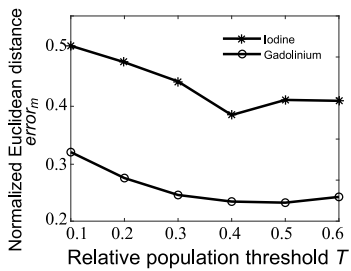}
	\caption{Influence of relative population threshold on the normalized Euclidean distance for iodine and gadolinium.}
	\label{figure.pt}
\end{figure}

Finally, the influence of  relative population threshold $T$ is illustrated in Fig. \ref{figure.pt}. Both too small or big $T$ leads to larger $error_{m}$. Nevertheless, the error varies relatively smoothly with T, especially for gadolinium. In our experiments, $T=0.4$ corresponds to the smallest  $error_{m}$.

\begin{table}
	\centering
	\caption{The normalized Euclidean distances $error_{m}$ of iodine with different parameters on physical phantom.}
	\begin{tabular}{c||cccc	}
		\hline
		{\diagbox{$\theta$}{$\sigma^2$}}&0.5&1&2&4\\
		\hline
		\hline	
		0  &	0.85&0.61&0.37&0.61				\\	
		0.2&	0.38&0.38&0.38&0.38				\\
		0.4&	0.39&0.39&0.39&0.39				\\
		0.6&	0.39&0.63&0.63&0.63				\\
		0.8&	0.40&0.63&0.63&0.40			\\
		1&		0.50&0.50&0.50&0.50			\\
		\hline
	\end{tabular}
	\label{Tab.error.parameters}
\end{table}

\section{Discussion}\label{sec.Discussion}
We have proposed a ROI-wise material decomposition method.
The proposed method enables materials to be more reliably and accurately decomposed.  This is mainly due to the introduction of spatio-spectral segmentation that allows pertinent features encoded in multi-energy sCT images to be extracted for basis materials optimization.

The results show that the ROI-wise material decomposition method favors more reliability while the sensitivity of detecting materials is somewhat sacrificed as the trade-off. 
This may partly explain why only materials with a concentration over certain limit could be accurately decomposed. 
This is the case for iodine with concentration 1 $mg/cc$ compared to higher concentrations (Fig. \ref{figure.decomposition.gray}). 
These results are consistent with the high improvement of FP (64\% improvement for iodine and 96\% for gadolinium compared to Coarse method on physical phantom) and worse performance of FN. 
Although Coarse and TV methods show better FN, they cannot separate materials of small concentrations or water-like materials. In other words, Coarse or TV method produced smaller FN at the cost of confusing water-like materials with materials of interest. For example, smaller FN of Coarse method compared to ROI method on physical phantom is ascribed to the fact that Coarse method was not able to detect gadolinium of 2 $mg/cc$ from water or iodine of 1 $mg/cc$.
The above detection limit of gadolinium and iodine for the proposed method is due to multiple factors, such as sCT image reconstruction quality and the performance of ROI segmentation.
Actually, the trade-off between sensitivity and reliability of the proposed ROI-wise method is regularized  by the threshold $T$ in the relative population thresholding. 
As illustrated in Fig. \ref{figure.pt}, too small or too large $T$ induced the increase of decomposition errors, because smaller $T$ leads to poor reliability (but high sensitivity). 
In other words, when $T$ is too small, noise and reconstruction errors will have a strong impact on decomposition accuracy. On the opposite, too large $T$ will degrade the decomposition ability for materials of small concentration.
Fortunately, the results show a relatively large range for the choice of $T$ around the optimal value, which implies that the proposed method is relatively little sensitive to $T$.

Finally, it is worth noting that image reconstruction quality has dramatical influence on the performance of material decomposition in image domain. 
Because of excessively low dose allocated to detectors, sCT reconstruction at each energy bin is a problem of low-dose CT reconstruction, which is also a challenging problem.  
A worse image reconstruction quality may deteriorate the performance of ROI segmentation, which in return may influence the precision of ROI-wise method, especially for images containing small structures ($e.g.$ small blood vessels).
We have chosen a common but efficient algorithm in the field of low-dose CT reconstruction (SART-TV) to reconstruct sCT images.
However, the results of reconstruction still show obvious artifacts and noise (Fig. \ref{fig.PhysicalPhantom.b}). Nevertheless, even in this situation, the materials were still correctly decomposed on both digital and physical phantoms, which demonstrates the robustness and reliability of the proposed ROI-wise method.  

\section{Conclusions}\label{sec.Conclusion}

We have proposed a ROI-wise material decomposition method for sCT by optimizing basis materials. This is achieved through spatio-energy segmentation and exploiting both morphological and spectral information in the sCT images. The results on digital and physical phantoms showed that the ROI-wise material decomposition method presents clearly higher accuracy and reliability compared to common decomposition methods based on TV or $lasso$ regularization. In the future work, the ability of detecting low-concentration materials will further be investigated to improve the sensitivity of the method while maintaining reliability.
Meanwhile, the ability of identifying small structures having small material concentration will also be investigated.

\section*{Acknowledgment}
We thank Dr. Cyril Mory for his help in preparing the physical phantom data and Dr. Simon Rit and Dr. Yoad Yagil for the careful reading of the manuscript and many helpful comments.

\bibliographystyle{IEEEtran}
\bibliography{spectral.bib}

\end{document}